# Early Indicators of Scientific Impact: Predicting Citations with Altmetrics

**Akhil Pandey Akella**
Department of Computer Science, Northern Illinois University, DeKalb, IL, USA
aakella@niu.edu

**Hamed Alhoori**
Department of Computer Science, Northern Illinois University, DeKalb, IL, USA,
alhoori@niu.edu

**Pavan Ravikanth Kondamudi**
Department of Computer Science, Northern Illinois University, DeKalb, IL, USA,
pavanravikanth.kprk@gmail.com

**Cole Freeman**
Department of Computer Science, Northern Illinois University, DeKalb, IL, USA,
cole.freeman9@gmail.com

**Haiming Zhou**
Department of Statistics and Actuarial Science, Northern Illinois University, DeKalb, IL, USA,
zhouh@niu.edu

**Abstract**
Identifying important scholarly literature at an early stage is vital to the academic research community and other stakeholders such as technology companies and government bodies. Due to the sheer amount of research published and the growth of ever-changing interdisciplinary areas, researchers need an efficient way to identify important scholarly work. The number of citations a given research publication has accrued has been used for this purpose, but these take time to occur and longer to accumulate. In this article, we use altmetrics to predict the short-term and long-term citations that a scholarly publication could receive. We build various classification and regression models and evaluate their performance, finding neural networks and ensemble models to perform best for these tasks. We also find that Mendeley readership is the most important factor in predicting the early citations, followed by other factors such as the academic status of the readers (e.g., student, postdoc, professor), followers on Twitter, online post length, author count, and the number of mentions on Twitter, Wikipedia, and across different countries.

**Keywords**: Citation Count, Citation Prediction; Altmetrics, Scientometrics, Scholarly Communication, Social Media, Science of Science, Scholarly Impact

## 1. Introduction

Every year an extraordinary volume of scientific literature is published. The number of scholarly articles in existence is doubling every nine years (Van Noorden, 2014)—a rate that is expected to increase with the proliferation of online publishing venues. Searching through this expanding body of work to find ideas that are new, important, and relevant to a given research direction has become challenging, even daunting, for many, if not all, researchers in the scientific community. Yet, many of the essential functions of academia depend on performing this task effectively. Researchers themselves benefit from up-to-date knowledge of their field, as discovering an article with new findings or surprising conclusions can influence the direction of their research projects. Academic institutions rely on accurate assessments of scholarly impact in their hiring practices, performance evaluations, and promotion decisions. Similarly, university rankings depend in part on similar methods of evaluating scholarly impact.

In response to the ever-increasing complexity and importance of identifying crucial research findings and assessing scholarly impact, a new multidisciplinary field known as the Science of Science has emerged (Zeng et al., 2017). Its fundamental purpose is to understand the mechanisms through which science is performed, received, and evaluated. Researchers in this field have endeavored to predict and analyze many aspects of scientific research. Some of these include predicting rising stars (van Dijk et al., 2014) or Nobel Prize winners (Gingras & Wallace, 2010; Revesz, 2014b), identifying scientific concepts that will gain traction (McKeown et al., 2016), ranking articles (Totti et al., 2016), and evaluating the long-term impact of citations on academic careers (D. Wang et al., 2013).

The research community has used many approaches to determine an article's scholarly impact, but peer review and citation analysis have been among the most important. In the last several decades, evaluating articles by the number of citations they generate has become the gold standard for scholarly appraisal. Yet, it can take months, even years, for an article to accumulate citations. As a complement to these traditional measures of scholarly impact, a new area known as altmetrics (Thelwall et al., 2013) has become a subject of interest. They capture the dissemination of research outcomes via multiple online platforms. The use of these platforms as a way to determine scholarly interest has some noteworthy advantages over traditional methods. Information is propagated online at a much faster rate than is the case with traditional citations. Online sources also provide access to a larger volume of information about research outcomes. These two factors provide the ideal conditions for the application of machine learning, which could provide quick insights into which research conclusions are likely to gain traction within a given field. Although not a substitute for traditional methods of scholarly evaluation, altmetrics have the advantage of making use of the broader spectrum of information available in the digital age.

Finding and adopting more efficient methods for discovering new and important articles would greatly increase the rate at which scientific ideas are absorbed. In the discovery period, researchers often sift through a high volume of articles to determine which ones are most relevant to their work, taking time away from the more creative aspects of research. This problem compounds year by year as more and more research articles proliferate across a growing number of online platforms. Social media has emerged as the primary informal channel of communication in the past decade, and researchers are increasingly using these platforms to communicate the results of their research and as a way to share articles, ideas, and evaluations of research with their peers. The main advantage of these new channels of communication is that gaining attention is a much faster process than with traditional citations. Major content begins to spread online within hours. Along with social media and online reference managers, research is often picked up online and mentioned in other online news outlets, blogs, and online peer review websites (Alhoori et al., 2018). Further, online and social media platforms have been used to predict several events related to education, health, business, and politics (Asur & Huberman, 2010). Using these new channels of communication to predict which articles will become important could greatly improve the discovery process, enabling researchers to identify new ideas relevant to their own projects more rapidly than is currently possible. In a broader sense—the sense that matters most to the scientific community and to the general public—by increasing the productivity rate of individual researchers, the process of scientific discovery will accelerate and the implementation of scientific results likewise.

In this study, we investigate altmetrics as a way to gain insight into the scholarly impact of research. We use features derived from social media shares, mentions, and reads of thousands of scientific articles across various platforms to train machine learning models to predict citation counts. The set of articles we use in our study represents a wide range of scientific disciplines, and we compare our predictions to the actual number of citations the articles received by the year or four years after we collected our altmetric data. We appraise our models, evaluating their accuracy in predicting citation counts with several metrics, and analyze which types of models are most useful for our data and why specific models are more effective than others. We present the results of our experiments to accelerate scientific discovery and improve the process by which researchers seek knowledge.

## 2. Related Work

The literature includes numerous studies focused on measuring and predicting the scholarly impact (Bai et al., 2016; Bornmann & Daniel, 2008; Kulkarni et al., 2007; Lokker et al., 2008; Onodera & Yoshikane, 2015; Penner et al., 2013; Price, 1976; Ruan et al., 2020; Waltman, 2016; M. Wang et al., 2012; Xiao et al., 2016). Early studies found a relationship between media coverage of scientific articles and scholarly impact (Chapman et al., 2007; Kiernan, 2003; Phillips et al., 1991). Several studies have attempted to predict the scholarly impact of scientific articles using early citations (Abramo et al., 2019; Abrishami & Aliakbary, 2019; Cao et al., 2016; Li & Tong, 2015; Mazloumian, 2012). Yan et al. (2011, 2012) considered a range of features to predict citations that include author, content, venue, venue rank, venue centrality, topic rank, diversity, h-index, author rank, productivity, sociality, and authority. Their evaluation of several predictive models found that authors of scholarly articles show signs of prejudice in their citation practices. The fundamental observation was that the features, author expertise and venue impact are crucial in determining a paper's citation count. Nezhadbiglari, Gonçalves, and Almeida (2016) computed the popularity trend using the spectral clustering algorithm based on the total citation count as a measure of scientific popularity. By extracting a set of academic features such as the number of publications a scholar has and the number of venues in which an author has published, the computed popularity trend was useful in predicting the early fame of a scholar. Weihs and Etzioni (2017) used 63 author and article features and machine learning techniques with probabilistic modeling approaches to predict scholarly citations and h-indices.

The h-index has been considered in a number of research articles in terms of its predictive power. Hirsch (2007) found the h-index to be an important feature in predicting future scholarly achievement. Acuna, Allesina, and Kording (2012) attempted to predict what the h-index of several authors in the medical field would be five years in the future. They collected information related to publications, funding, and citations for 3,085 neuroscientists, 57 Drosophila researchers, and 151 evolutionary scientists. They used regression models to predict the researchers' h-indices and found the most important predictive factors to be number of articles published, diversity of articles in distinct journals, and number of articles published in prestigious journals. In investigating the problem of predicting an author's h-index, Dong et al. (2015, 2016) explored features related to content, including authors, publishing venues, and social and temporal data. They found the author's authority on the publication topic as indicated by numerous citations of their work by experts in the area as well as the publication venue to be the most influential factors.

Castillo, Donato, and Gionis (2007) predicted citations using author-based features, link-based features, and early citations. Similarly, Fu and Aliferis (2008) used author, venue, and citation-related features to predict the number of citations an article would have ten years after its publication. They used thresholds of 20, 50, 100, and 500 citations and achieved an area under the receiver operator curve (AUC) between 0.857 and 0.918, indicating that their predictions achieved high precision and recall scores. Chakraborty, Kumar, Goyal, Ganguly, and Mukherjee (2014) used author, venue, and article features to classify citations into six categories and then built a regression model to predict the citation count within each category. T. Yu, Yu, Li, and Wang (2014) used regression analysis to predict citations by considering features related to an article, author, citations, and publication journal. Chen and Zhang (2015) performed a regression analysis and built machine learning models to predict the citation count of research articles. They considered a set of six content features and ten author features in building the regression models. Stegehuis, Litvak, and Waltman (2015) proposed quantile-based regression models to predict future citations. Their models performed best when two variables (impact factor and early citation counts) were used together instead of separately. Sarkar, Lakdawala, and Datta (2017) investigated a citation prediction of topics in software engineering and achieved high prediction accuracy.

Social factors that may influence citation count are the focus of another line of research (Nicolaisen, 2003). For example, higher centrality in the coauthorship network (Sarigöl et al., 2014) and a higher level of collaboration (Figg et al., 2006; Katz & Hicks, 1997; Wuchty et al., 2007) were found to increase the number of citations. Manjunatha, Sivaramakrishnan, Pandey, and Murthy (2003) and Davletov, Aydin, and Cakmak (2014) used various machine learning techniques to consider temporal features in predicting the citation performance of articles. In addition to machine learning models, statistical techniques have also been applied; for example, the ordered probit model, a statistical model used by (Perlich, Provost, and Macskassy, 2003), and a spatio-temporal function (Revesz, 2014a) have all been used to predict an article's citation count. Badache and Boughanem (2017) used social signals to rank the importance of articles based on a consideration of temporal aspects.

Social media data have also been considered in predicting citations (Finch et al., 2017; Wooldridge & King, 2019). Kwak and Lee (2014) studied scholarly sharing in Twitter communities and found that measuring scientific impact through the lens of social media to be of limited value given the focus on just a few top journals. Using binary classification, Zoller, Doerfel, Jäschke, Stumme, and Hotho (2016) drew on logs from the BibSonomy social tagging system and found that bookmarks, exports, and views of publications have mild correlations with citations from Microsoft Academic and are useful for predicting future citations. Using machine learning models and social media metrics, Kale, Siravuri, Alhoori, and Papka (2017) investigated whether or not research articles would be cited in policy documents and achieved an accuracy rate of 87%. Thelwall and Nevill (2018) used Scopus citations and built a linear regression model to predict citations using altmetrics. Recently, Lehane and Black (2020) found a positive correlation between citations and altmetrics in the field of critical care medicine.

Graph mining techniques and link prediction have also been used to predict citation counts (Bütün et al., 2017; Pobiedina & Ichise, 2016, 2014; Sebastian, 2014; Timilsina et al., 2017; X. Yu et al., 2012). In a study on the joint modeling of texts, Nallapati, Ahmed, Xing, and Cohen (2008) analyzed the Pairwise-Link-LDA and the Link-PLSA-LDA models for citation prediction. Of the two models, the researchers observed that the Link-PLSA-LDA model performed better on the citation prediction task. Kunegis, Fay, and Bauckhage (2010) proposed a link prediction algorithm based on a spectral evolution model, according to which the growth can be described by a change in the spectrum. They studied the comparison of the graph kernel function (a variety of link prediction algorithms) and found that the spectral evolution model provides a justification for more complex link prediction methods. Sun, Han, Aggarwal, and Chawla (2012) used a meta-path model with a time-prediction model to not only predict citations but also the timeframe within which they would accrue. Shibata, Kajikawa, and Sakata (2012) examined link prediction on five large citation network datasets and observed F-1 scores of 0.74 and 0.82. They found out that the topological features are very important for link prediction in citation networks.

Research articles have different levels of influence on cited or citing references (Dietz et al., 2007). Researchers cite research articles for different reasons. Even within a given article, researchers focus on different components of a cited article if they cite it multiple times (Elkiss et al., 2008). Some studies follow this line of thought by exploring citation classification (Teufel et al., 2006). Zhu, Turney, Lemire, and Vellino (2015) proposed a machine learning approach using various features to identify the cited articles with the greatest influence on a given publication. They found that the number of times an article is cited in the body of the article is the most important feature for this classification task. Valenzuela, Ha, and Etzioni (2015) used 465 annotated citations to predict whether a scholarly citation is important to a cited work or not using NLP and supervised classification. Hassan, Akram, and Haddawy (2017) expanded the previous study by adding six new features and achieved a higher accuracy.

Sinatra, Wang, Deville, Song, and Barabási (2016) studied the evolution of researchers' scholarly outcomes and found that the publications with the highest impact are randomly distributed across researchers' careers (e.g., first, middle, or last publication). Singh et al. (2017) analyzed the effects of various types of citations at different points in the life cycle of an article. They compared the effects of "influential" and "non-influential" authors citing a work early in the life cycle. They also compared the effects of early self-citations, co-author citations, and citations by more distant authors. They found a negative correlation between early citations by high-impact authors and long-term citation count.

Many other factors have been explored in relation to citation count. These include open access research articles (Alhoori et al., 2015), first research articles published in an area (Newman, 2009), time since publication (Burrell, 2002), subject of study, study design, and document properties (Didegah & Thelwall, 2013b), journal impact (Didegah & Thelwall, 2013a), textual features (Ibáñez et al., 2009), number of downloads (Brody et al., 2006; Harnad & Brody, 2004), tweets (Eysenbach, 2011; Peoples et al., 2016; Shuai et al., 2012; Tonia et al., 2016), Facebook likes (Ringelhan et al., 2015), geographic location (Bornmann & Leydesdorff, 2012), funding sources and characteristics of the title and article (Sagi & Yechiam, 2008), citation sentiments (Kumar, 2016), newsworthiness (Callaham et al., 2002; Siravuri & Alhoori, 2017), citation context (Moed, 2010; Singh et al., 2015), and number and quality of references (Antoniou et al., 2015; Jiang et al., 2013; Tahamtan et al., 2016).

Overall, the literature shows that in efforts to predict the scholarly impact of research articles, researchers have considered many approaches and features. The literature also shows that for most of the indicators considered, determining impact is a waiting game given the lack of data for newly published articles. That is, significant time must elapse before a research article either accumulates citations or it becomes evident that the article has not and, therefore, is unlikely to accumulate them. In the present study, we address this problem using a variety of models and features based on altmetrics.

## 3. Data

We collected social media mentions of research articles from Altmetric.com. The dataset consisted of altmetrics for more than 5 million articles, from which we selected a random sample of 12,374 articles that were published in 2015. Figure 1 shows the distribution of the articles by their major discipline. As we were interested in determining the extent to which altmetric features can serve as a complement to traditional measures of scholarly impact, we used citation counts for these articles as the target variable for our models. We searched Google Scholar using the DOI of each article and collected the citation count with a custom scraping tool.

Our experiments test whether the response to articles on social media, which occur relatively quickly after publication, can serve as an accurate predictor of the citation counts those will achieve after a longer period of time. To do this, it was necessary to separate the point at which we recorded the altmetric features from the point at which we determined the citation counts for each article. Therefore, we collected our altmetric records for each article in June 2016 and the citation counts in September 2017 and October 2020, which gave us a difference of more than a year between our predictors and the responses in both cases (2017 and 2020). We used the citations from 2017 as the target variable for the main experiment. We repeated and adapted the same experimental structure for the 2020 citations. The long-term citations helped us analyze our inference in relation to our results for the early citations.

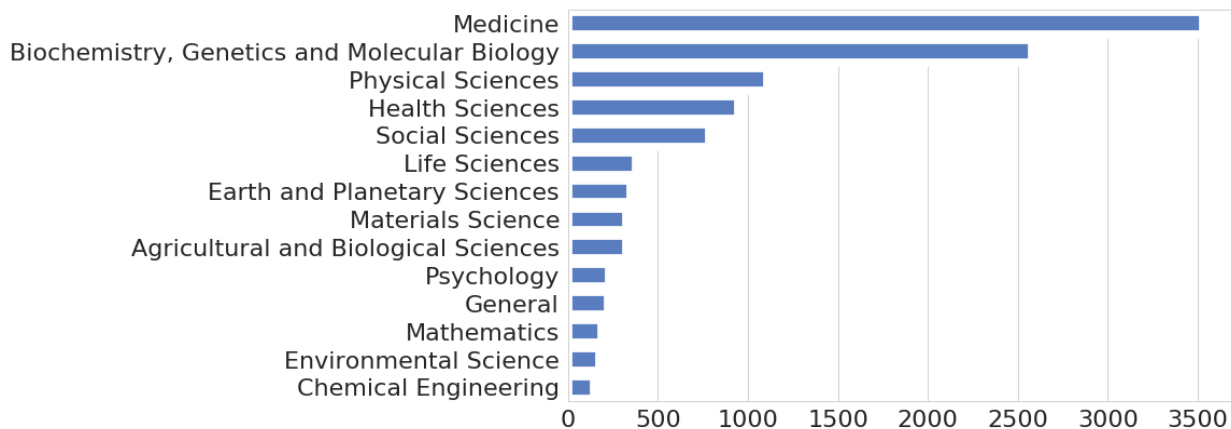

**Figure 1:** Distribution of articles by discipline.

We collected a total of 21 features, as listed and described in Table 1. From those features, 18 are quantitative, representing readerships on online reference managers, such as Mendeley, as well as counts of the number of posts (e.g., tweets, Facebook shares) on social media platforms. Three of the features—academic status, profession on Twitter, and platform with maximum mentions—are qualitative. Table 2 shows the descriptive statistics of the features.

**Table 1**. Principal features of our data.

| Feature | Feature description |
|---|---|
| Author Count | Number of authors of a research article |
| Mendeley Readership | Number of times an article has been referenced on Mendeley |
| CiteULike Readership | Number of times an article has been bookmarked on CiteULike |
| News Mentions | Number of times an article has been mentioned in news outlets |
| Blogs | Number of times an article has been mentioned or featured in blogs |
| Reddit | Number of times an article has been mentioned on Reddit |
| Twitter | Number of times an article has been mentioned in a tweet on Twitter |
| Retweets | Number of users who have retweeted an article |
| Twitter Mentions | Number of Twitter users mentioned in all tweets related to a given article (e.g., tweet = "tweet text @user1 @user2" = > Two users) |
| Facebook | Number of times an article has been shared on Facebook |
| GooglePlus | Number of times an article has been shared on Google Plus |
| Peer Review | Number of times an article has been shared on the major peer review websites Publons and PubPeer |
| Wikipedia | Number of times an article has been cited on Wikipedia |
| Total Platforms | Number of online platforms on which an article has been shared |
| Countries | Number of countries in which an article has been shared online |
| Max. Followers | Follower count of the user with the highest number of followers among all users who have tweeted or retweeted about an article (we included this user on the premise that he/she is likely to be more influential than other users) |
| Academic Status | Seniority or academic status of a user who has bookmarked a publication on Mendeley (examples: student, librarian, postdoc, researcher, and professor) |
| Profession on Twitter | Profession of a user who has tweeted about a publication (examples: practitioner, science communicator, and "unknown") |
| Platform with Max Mentions | Social media platform with the highest number of posts about an article |
| HashTags | Number of hashtags used by users in posts extracted from Twitter |
| Post Length | Total sum of the length of all posts related to an article across all platforms. Users often post just the link to and the title of an article. This redundant information was removed before post length was counted |

**Table 2.** Descriptive statistics of the features.

| Feature | mean | std | min | 25% | 50% | 75% | max |
|---|---|---|---|---|---|---|---|
| Author Count | 2.08 | 20.73 | 0 | 0 | 0 | 2 | 2245 |
| Mendeley Readership | 14.98 | 37.89 | 0 | 0 | 4 | 14 | 1042 |
| CiteULike Readership | 0.14 | 0.83 | 0 | 0 | 0 | 0 | 32 |
| News Mentions | 0.29 | 2.85 | 0 | 0 | 0 | 0 | 150 |
| Blogs | 0.15 | 0.86 | 0 | 0 | 0 | 0 | 51 |
| Reddit | 0.02 | 0.21 | 0 | 0 | 0 | 0 | 11 |
| Twitter | 4.58 | 26.00 | 0 | 1 | 1 | 3 | 1182 |
| Retweets | 2.22 | 19.16 | 0 | 0 | 0 | 1 | 920 |
| Twitter Mentions | 0.81 | 3.68 | 0 | 0 | 0 | 1 | 158 |
| Facebook | 0.46 | 8.54 | 0 | 0 | 0 | 0 | 893 |
| GooglePlus | 0.07 | 1.20 | 0 | 0 | 0 | 0 | 65 |
| Peer Review | 0.01 | 0.15 | 0 | 0 | 0 | 0 | 13 |
| Wikipedia | 0.15 | 0.48 | 0 | 0 | 0 | 0 | 9 |
| Total Platforms | 9.03 | 3.17 | 0 | 7 | 7 | 13 | 16 |
| Countries | 2.48 | 3.81 | 0 | 0 | 1 | 3 | 107 |
| Max. Followers | 8261.33 | 60014.82 | 0 | 4 | 509 | 2505 | 2406790 |
| HashTags | 0.93 | 2.98 | 0 | 0 | 0 | 1 | 83 |
| Post Length | 123.33 | 69.22 | 0 | 69.25 | 130 | 144 | 276 |

Multicollinearity (i.e., high correlations among features) can be problematic when estimating regression models as it leads to unstable estimates of regression coefficients and makes the estimated coefficients difficult to interpret. The interpretation of a regression coefficient usually involves gauging the effect that changes in one independent variable have on a dependent variable while holding other independent variables constant. However, if independent variables are highly correlated, it is unclear as to which is motivating changes in the dependent variable. It is, therefore, standard practice to remove one pair of highly correlated features before using data to build models. Figure 2 shows the correlation matrix for the features in our dataset. The highest correlations are between Twitter and Retweets. Because of this high collinearity, we removed the Retweets.

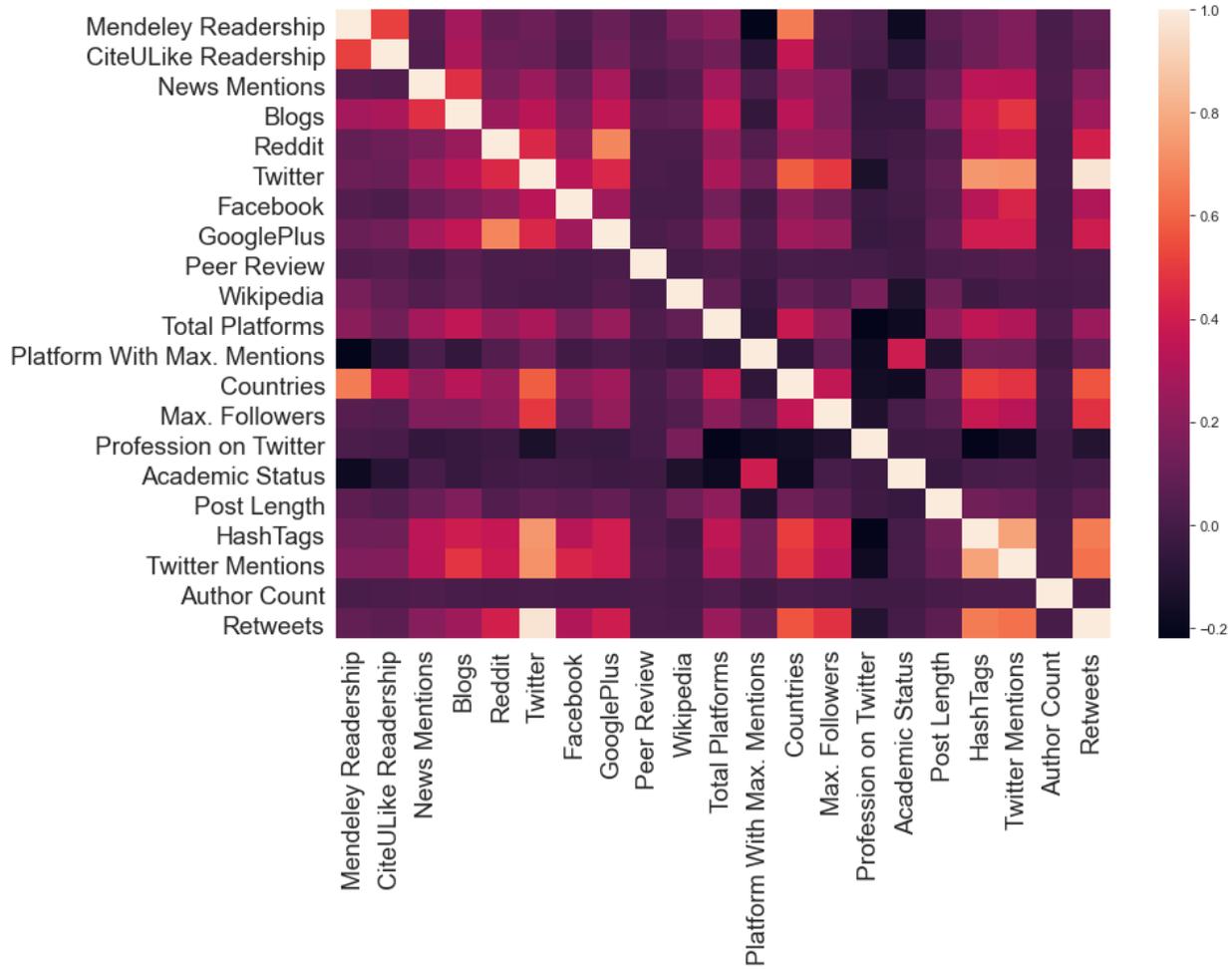

**Figure 2**: Correlation matrix between features.

## 4. Methods
### *4.1 Experiments*
With this data, we devised a set of three experiments to test the relationships between altmetric features and the scholarly attention that the articles received (i.e., eventual citation count). In each experiment, we used the same features to build machine learning models of different types to estimate a unique target variable. The goal of our model-building process was to identify the optimal parameters for each type of model in a given circumstance. When we had achieved optimal performance for our models, we evaluated their relative success in predicting a given target with several measures of accuracy. We also analyzed the most successful models to determine which altmetric features are the best predictors of scholarly impact.

In our first two experiments, we focused on binary classification problems. The goal of the first experiment was to predict whether a given article would receive (i) zero or (ii) one or more citations. The goal of the second experiment was to predict whether an article would receive (i) more or (ii) fewer than the median number of citations of all articles in our set. In the third experiment, we built regression models to predict the number of citations a given research article will receive.

To determine which features are most important for yielding accurate predictions, we used information gain (i.e., entropy) ((Shannon, 1948) and Gini impurity (Breiman et al., 1984) to represent the contribution of each feature. We used four metrics to evaluate the results of all three experiments: accuracy, precision, recall, and F-1 scores. Table 3 shows the confusion matrix, which is a special contingency table with two dimensions: *actual* and *predicted*. The categories represented in the confusion matrix each factor into the accuracy metrics in a different way. Each metric provides an important measure of how a given model performed based on specific criteria.

Table 3. Confusion matrix.

| | | Predicted class | |
|---|---|---|---|
| **Actual class** | | Positive | Negative |
| | **Positive** | True Positive (TP) | False Negative (FN) |
| | **Negative** | False Positive (FP) | True Negative (TN) |

Model accuracy is the percentage of samples correctly classified by the model, as shown in Eq. (1):

$$Accuracy = \frac{TP+TN}{TP+FN+FP+TN} \qquad (1)$$

Model precision is the percentage of results the model predicted as positive that are actually positive, as shown in Eq. (2):

$$Precision = \frac{TP}{TP+FP} \qquad (2)$$

Model recall is the percentage of relevant results retrieved, as shown in Eq. (3):

$$Recall = \frac{TP}{TP+FN} \qquad (3)$$

The F-1 score represents the harmonic mean of the precision and recall scores. It is computed using Eq. (4). All measures of accuracy are bounded in the range [0,1], with 0 representing the lowest possible score and 1 the highest:

$$F1 = 2(\frac{Precision \times Recall}{Precision + Recall}) \qquad (4)$$

*4.2 Model construction*
For our neural networks, we used a fully connected deep neural network. To obtain the appropriate combination of hyperparameters, we experimented with different permutations of values. These hyperparameters include the batch size, the number of epochs, the number of hidden layers, and the number of neurons required for each hidden layer, and the optimization, loss, and activation functions. We chose those that performed best, including one hidden layer with 512 neurons and SELU (Klambauer et al., 2017) as the activation function for the hidden layer. As this was a classification problem, we selected Softmax (Bridle, 1990) as the activation function for the output layer. We trained the network for 100 epochs with a batch size of 32 using RMSprop (Tieleman & Hinton, 2012) as the optimization function. The learning rate for the activation function was 0.001. The cost function for the network was binary cross-entropy. We implemented the neural network using TensorFlow (Abadi et al., 2016).

We also trained and tested some supervised learning algorithms using scikit-learn implementations (Pedregosa et al., 2011), including the Random Forest, Decision Tree, Gradient Boosting, AdaBoost, Bernoulli Naive Bayes, and K-Nearest Neighbors (KNN) classifiers. All these algorithms were trained using 10-fold cross-validation. These algorithms have several advantages and disadvantages (Han et al., 2011; Hastie et al., 2003; Tan et al., 2019). For example, an ensemble method (Dietterich, 2000) combines a group of base classifiers. As a result, it often produces relatively high prediction accuracy and is robust to errors and outliers. However, ensemble methods are more computationally expensive than other models such as the Decision Tree, and the results can be more difficult to interpret due to the variety of the base classifiers that are combined to produce class predictions. Finally, we used the C-support vector classification algorithm (Chang & Lin, 2011) and discovered the optimal parameters for our problem using a randomized search.

We performed hyperparameter tuning in order to identify the best set of tuning parameters to train the machine learning models. We achieved this by using a randomized search and a grid search. We decided to perform parameter tuning on specific models where optimization has been shown to improve results most significantly: Random Forest, Decision Tree, and Gradient Boosting in Experiment 1 and Experiment 2. The remaining three classification algorithms were trained using the default hyperparameters provided by scikit-learn. For Experiment 3, we performed hyperparameter tuning on only the Random Forest and Decision Tree models. The parameters used for hyperparameter tuning are as follows:

- **Number of trees in the forest (num estimators)**: Number of trees produced to estimate the samples.
- **Minimum samples required to split a node (min samples split)**: Smallest number of samples required to split an internal node (increasing this parameter causes the tree to store more samples at each node).
- **Minimum samples required to be at a leaf node (min samples leaf)**: Minimum number of samples required at a leaf node.
- **Maximum features (max features)**: Maximum number of features each tree should consider to perform the best split.
- **Maximum depth of the tree (max depth):** Maximum depth of the tree the model can use to capture information about the samples (the maximum depth of the tree is directly proportional to the number of splits, which is indicative of how much information is collected).
- **Criterion**: Criteria required to measure the quality of a split (i.e., Gini impurity and information gain).
- **Learning rate:** Impact of every tree on the final outcome for a sample (it controls the degree of change in estimating the sample).

For SVM, we tried linear and sigmoid kernels, with sigmoid emerging as the better option. The following are SVM tuning parameters for Experiments 1 and 2: kernel = "sigmoid," degree of kernel = 3, tolerance = 0.001, and gamma = 0.045.

## 5. Results
### 5.1 Experiment 1

In Experiment 1, we tested social media data to determine if it is useful for predicting whether or not an article will receive at least one citation. We considered this as a classification problem and encoded our predictor variable with binary class labels "YES" and "NO." The dataset comprised 9,779 observations belonging to the YES class and 2,595 observations belonging to the NO class. We split the data according to an 80:20 ratio for training and testing, respectively. To solve the first two classification problems, we used Neural Networks, Decision Tree, Bernoulli Naive Bayes, K-nearest Neighbors, SVM, and the ensemble algorithms Gradient Boosting, AdaBoost, and Random Forest. The accuracy, precision, recall, and F-1 results are shown in Table 4 for the citations from 2017. Of the models we built, AdaBoost performed best, slightly improving on the performance of Random Forest, Decision Tree, Gradient Boosting, and SVM. For the 2020 citation predictions, all the models achieved a higher F1 score than for the citations from 2017, as shown in Table 5. Table 6 shows the optimum tuning parameters. Figure 3 shows the features arranged from most to least important for the top-performing algorithms for 2017 citations. We ranked the features in regard to their importance to the Decision Tree and Random Forest classifiers from most to least important based on the information gain. We can see that Maximum Followers on Twitter, Post Length, and Mendeley Readership are the most important features.

**Table 4**. Comparison of model performance metrics for eight classifiers in Experiment 1 for predicting the 2017 citations.

| Model | Accuracy | Precision | Recall | F-1 |
|---|---|---|---|---|
| **Random Forest** | 0.783 | 0.783 | 1.0 | 0.878 |
| **Decision Tree** | 0.783 | 0.783 | 1.0 | 0.878 |
| **Gradient Boosting** | 0.783 | 0.783 | 1.0 | 0.878 |
| **AdaBoost** | **0.793** | 0.811 | 0.960 | **0.879** |
| **Bernoulli Naive Bayes** | 0.779 | 0.861 | 0.856 | 0.858 |
| **KNN** | 0.760 | 0.806 | 0.912 | 0.856 |
| **Neural Network** | 0.787 | 0.787 | 0.787 | 0.787 |
| **SVM** | 0.782 | 0.783 | 0.997 | 0.877 |

Table 5. Comparison of model performance metrics for eight classifiers in Experiment 1 for predicting the 2020 citations.

| Model | Accuracy | Precision | Recall | F-1 |
|---|---|---|---|---|
| **Random Forest** | 0.959 | 0.959 | 1.0 | 0.979 |
| **Decision Tree** | 0.959 | 0.959 | 1.0 | 0.979 |
| **Gradient Boosting** | 0.959 | 0.959 | 1.0 | 0.979 |
| **AdaBoost** | 0.958 | 0.959 | 0.999 | 0.979 |
| **Bernoulli Naive Bayes** | 0.959 | 0.959 | 1.0 | 0.979 |
| **KNN** | 0.959 | 0.959 | 1.0 | 0.979 |
| **Neural Network** | 0.790 | 0.790 | 0.790 | 0.790 |
| **SVM** | 0.956 | 0.959 | 0.997 | 0.978 |

Table 6. Optimum tuning parameters in Experiment 1.

|  | Random Forest | | Decision Tree | | Gradient Boosting | |
|---|---|---|---|---|---|---|
|  | 2017 Citations | 2020 Citations | 2017 Citations | 2020 Citations | 2017 Citations | 2020 Citations |
| **Random State** | 4051 | 377 | 2564 | 1313 | 3766 | 3703 |
| **Num estimators** | 32 | 2 | X | X | 2 | 32 |
| **Min samples split** | 0.9 | 0.4 | 1.0 | 0.6 | 1.0 | 0.8 |
| **Min samples leaf** | 0.1 | 0.5 | 0.1 | 0.4 | 0.1 | 0.2 |
| **Max features** | 12 | 3 | 16 | 14 | 8 | 4 |
| **Max depth** | 31 | 16 | 32 | 1 | 16 | 18 |
| **Criterion** | Entropy | Entropy | Gini index | Entropy | X | X |
| **Learning rate** | X | X | X | X | 0.01 | 0.5 |

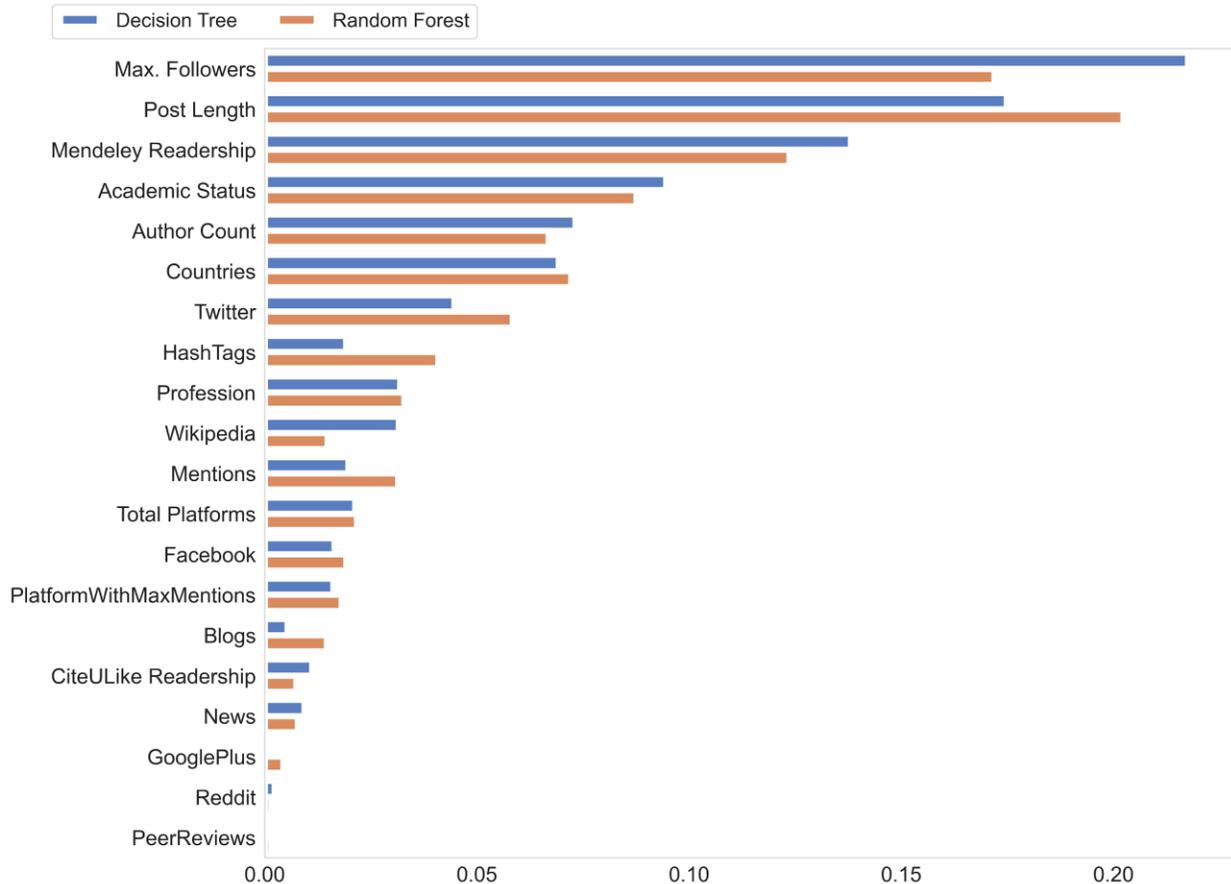

**Figure 3**: The most important features for Decision Tree and Random Forest classifiers for the 2017 citations in Experiment 1.

## 5.2 Experiment 2

In Experiment 2, we examined whether the altmetric features could be used to predict whether or not a publication will receive more than the median number of citations, which in our set of articles was eight. We labeled our publications using the binary indicators: YES, which means that an article received more than eight citations, and NO, which means the article received eight or fewer citations. In this experiment, the classes are almost equally balanced with 5,854 YES indicators and 6,520 NO indicators. We performed another binary classification as in Experiment 1.

The accuracy, precision, recall, and F-1 scores are shown in Table 7. Whereas the results from Experiment 1 were relatively close together in almost all the categories of accuracy measure and model type, the scores for Experiment 2 showed greater variation. Neural Networks, AdaBoost, and Gradient Boosting performed best by F-1 score. SVM achieved the lowest performance by all measures. The preferred model, however, depends on the specific task a researcher is pursuing. For example, if a model that minimizes false-positive cases (i.e., articles predicted to receive more than the median but failed to do so) is required, model performance would be judged by precision. In this case, Random Forest would be the model of choice. Table 8 shows the predictions for the 2020 citations and that all the models performed to a low level in this regard. Table 9 shows the optimum tuning parameters, and Figure 4 shows the feature importance of Gradient Boosting and AdaBoost. Mendeley Readership, Maximum Followers on Twitter, Academic Status, Countries, Author Count, and Post Length are the most important features for these classifiers.

**Table 7**. Comparison of model performance metrics for eight classifiers in Experiment 2 for predicting the 2017 citations.

| Model | Accuracy | Precision | Recall | F-1 |
|---|---|---|---|---|
| **Random Forest** | 0.749 | **0.807** | 0.646 | 0.718 |
| **Decision Tree** | 0.752 | 0.794 | 0.671 | 0.727 |
| **Gradient Boosting** | 0.772 | 0.789 | 0.734 | 0.760 |
| **AdaBoost** | 0.775 | 0.802 | 0.721 | 0.760 |
| **Bernoulli Naive Bayes** | 0.664 | 0.748 | 0.481 | 0.586 |
| **KNN** | 0.675 | 0.695 | 0.607 | 0.648 |
| **Neural Network** | **0.769** | 0.769 | **0.769** | **0.769** |
| **SVM** | 0.595 | 0.680 | 0.337 | 0.451 |

**Table 8**. Comparison of model performance metrics for eight classifiers in Experiment 2 for predicting the 2020 citations.

| Model | Accuracy | Precision | Recall | F-1 |
|---|---|---|---|---|
| **Random Forest** | 0.502 | 0.500 | 0.295 | 0.371 |
| **Decision Tree** | 0.487 | 0.484 | 0.468 | 0.476 |
| **Gradient Boosting** | 0.501 | 0.497 | 0.208 | 0.293 |
| **AdaBoost** | 0.484 | 0.476 | 0.357 | 0.408 |
| **Bernoulli Naive Bayes** | 0.482 | 0.473 | 0.354 | 0.405 |
| **KNN** | 0.513 | 0.511 | 0.515 | 0.513 |
| **Neural Network** | 0.511 | 0.511 | 0.511 | 0.511 |
| **SVM** | 0.516 | 0.526 | 0.276 | 0.362 |

**Table 9.** Optimum tuning parameters in Experiment 2.

|  | **Random Forest** | | **Decision Tree** | | **Gradient Boosting** | |
|---|---|---|---|---|---|---|
|  | 2017 Citations | 2020 Citations | 2017 Citations | 2020 Citations | 2017 Citations | 2020 Citations |
| **Random State** | 3878 | 1750 | 435 | 3010 | 1985 | 1062 |
| **Num estimators** | 200 | 32 | X | X | 200 | 16 |
| **Min samples split** | 0.3 | 0.3 | 0.5 | 0.1 | 0.3 | 0.8 |
| **Min samples leaf** | 0.2 | 0.1 | 0.4 | 0.2 | 0.1 | 0.2 |
| **Max features** | 8 | 18 | 13 | 13 | 5 | 19 |
| **Max depth** | 31 | 8 | 1 | 32 | 15 | 12 |
| **Criterion** | Gini Index | Entropy | Entropy | Gini Index | X | X |
| **Learning rate** | X | X | X | X | 0.1 | 0.05 |

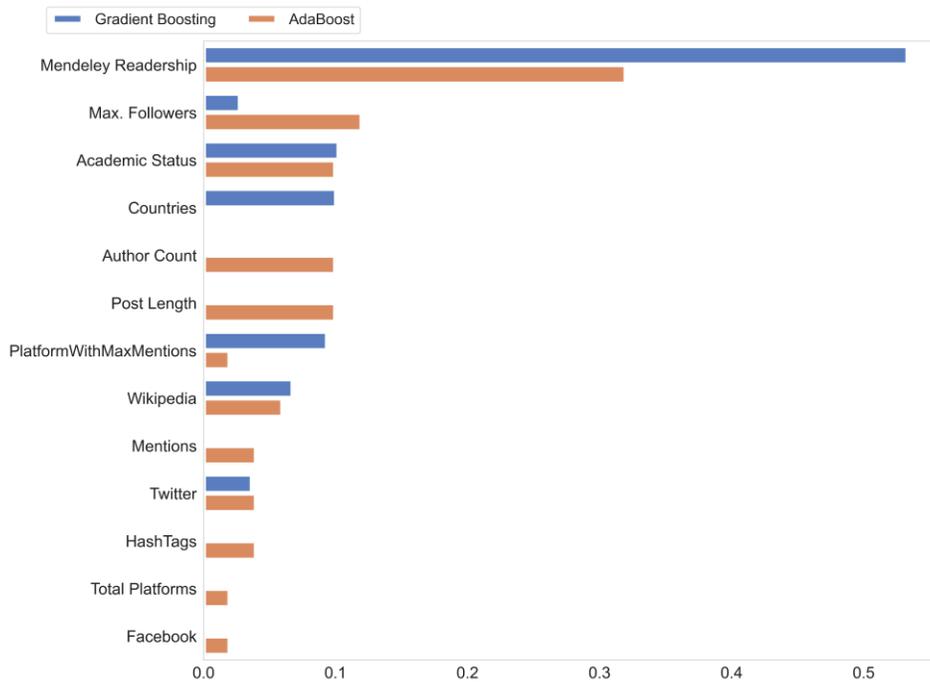

**Figure 4**: The most important features for Gradient Boosting and AdaBoost classifiers for the 2017 citations in Experiment 2.

### 5.3 Experiment 3

In Experiment 3, we considered fitting regression models whose response variable can be a continuous variable. By fitting regression models to the data, we were able to examine the factors that contribute to the prediction more precisely. For this experiment, we examined the performance of Multiple Linear Regression, Neural Network, and tree-based regression models (i.e., Decision Trees and Random Forest regressors). Before fitting the regression models, we performed a transformation to the response variable. We transformed the citation counts using the natural log function; i.e., the transformed variable is $\ln(1 + O)$, where O represents the original citation counts (Thelwall, 2016). In our Multiple Linear Regression model, we applied the same transformation to the other predictor variables—namely, Mendeley Readership, Wikipedia, Twitter, Max. Followers, Countries, Facebook, Twitter Mentions, CiteULike, HashTags, Blogs, GooglePlus, News, Reddit, Peer Reviews, and Author Count.

We calculated the R-squared and mean squared error (MSE) and the mean absolute error (MAE) on test data to compare all the models. R-squared measures the proportion of total variance in the number of citations explained by the model. MSE and MAE measure the predictive accuracy of the model. Although large R-squared values are a good sign of the model's ability to explain the variance in the target variable, they do not necessarily indicate model significance. This is also the case for MSE and MAE values. Table 10 shows the MSE, MAE, and R-squared values for the models for the 2017 citations. The Multiple Linear Regression model had the best MSE, MAE, and R-squared values and, therefore, performed better than the other models. Table 11 shows the values for the 2020 citations. Table 12 shows the optimum tuning parameters for the Random Forest and Decision Tree models for the 2017 citations. Table 13 shows the coefficients of the Multiple Linear Model for predicting the 2017 citations, whereas Table 14 shows the coefficients for the model predicting the 2020 citations. For the Random Forest regressor, the best criterion for measuring the quality of a split was the MAE (Willmott & Matsuura, 2005). Figure 5 shows the feature importance of the tree-based models and indicates that Mendeley Readership is the most important feature in predicting citations.

**Table 10**. Comparison of model performance metrics for four regressors in Experiment 3 for predicting the 2017 citations.

| Model | MSE | MAE | R-squared |
|---|---|---|---|
| **Random Forest** | 1.96 | 0.95 | 0.311 |
| **Decision Tree** | 1.61 | 0.98 | 0.395 |
| **Multiple Linear Model** | **1.50** | **0.94** | **0.435** |
| **Neural Network** | 1.91 | 0.98 | 0.292 |

**Table 11**. Comparison of model performance metrics for four regressors in Experiment 3 for predicting the 2020 citations.

| Model | MSE | MAE | R-squared |
|---|---|---|---|
| **Random Forest** | 1.25 | 0.87 | -0.001 |
| **Decision Tree** | 1.25 | 0.87 | -0.001 |
| **Multiple Linear Model** | 1.25 | 0.87 | 0.0008 |
| **Neural Network** | 1.29 | 0.89 | -0.05 |

**Table 12.** Optimum tuning parameters in Experiment 3 for predicting the 2017 citations.

| | Random Forest | | Decision Tree | |
|---|---|---|---|---|
| | 2017 Citations | 2020 Citations | 2017 Citations | 2020 Citations |
| **Num estimators** | 8 | 100 | X | X |
| **Min samples split** | 0.4 | 0.5 | 0.5 | 0.9 |
| **Min samples leaf** | 0.2 | 0.4 | 0.1 | 0.1 |
| **Max features** | 16 | 8 | 17 | 5 |
| **Max depth** | 4 | 9 | 32 | 32 |
| **Criterion** | MAE | MSE | Friedman MSE | Friedman MSE |
| **Random state** | 4944 | 4421 | 4927 | 4792 |

**Table 13.** Coefficients of the Multiple Linear Regression model for predicting the 2017 citations.

| | **Coefficient** | **Standard Error** | **t-value** |
|---|---|---|---|
| **Constant** | 1.2798 | 0.109 | 11.744 |
| **Mendeley Readership** | 0.6241 | 0.017 | 37.676 |
| **CiteULike Readership** | 0.2416 | 0.051 | 4.760 |
| **News Mentions** | 0.1149 | 0.041 | 2.809 |
| **Blogs** | 0.1542 | 0.054 | 2.834 |
| **Reddit** | -0.0374 | 0.147 | -0.255 |
| **Twitter** | -0.2385 | 0.033 | -7.215 |
| **Facebook** | -0.0801 | 0.038 | -2.102 |
| **GooglePlus** | -0.0150 | 0.078 | -0.193 |
| **Peer Review** | 0.1498 | 0.224 | 0.667 |
| **Wikipedia** | 0.9265 | 0.055 | 16.848 |
| **Total Platforms** | -0.0243 | 0.018 | -1.377 |
| **Platform with Max Mentions** | 0.0304 | 0.006 | 5.458 |
| **Countries** | 0.1332 | 0.029 | 4.644 |
| **Max. Followers** | -0.0455 | 0.006 | -7.226 |
| **Profession on Twitter** | -0.0077 | 0.016 | -0.488 |
| **Academic Status** | 0.0022 | 0.002 | 1.055 |
| **Post Length** | 0.0006 | 0.000 | 2.958 |
| **HashTags** | -0.0655 | 0.026 | -2.489 |
| **Twitter Mentions** | 0.2296 | 0.032 | 7.158 |
| **Author Count** | -0.0267 | 0.014 | -1.841 |

**Table 14.** Coefficients of the Multiple Linear Regression model for predicting the 2020 citations.

| | **Coefficient** | **Standard Error** | **t-value** |
|---|---|---|---|
| **Constant** | 2.8236 | 0.101 | 28.093 |
| **Mendeley Readership** | 0.0003 | 0.015 | 0.018 |
| **CiteULike Readership** | 0.0062 | 0.047 | 0.133 |
| **News Mentions** | -0.0491 | 0.038 | -1.302 |
| **Blogs** | 0.0316 | 0.050 | 0.629 |
| **Reddit** | -0.0456 | 0.136 | -0.336 |
| **Twitter** | 0.0047 | 0.030 | 0.153 |
| **Facebook** | -0.0473 | 0.035 | -1.347 |
| **GooglePlus** | 0.0113 | 0.072 | 0.158 |
| **Peer Review** | -0.2774 | 0.207 | -1.340 |
| **Wikipedia** | -0.1088 | 0.051 | -2.145 |
| **Total Platforms** | 0.0132 | 0.016 | 0.811 |
| **Platform with Max** | 0.0058 | 0.005 | 1.130 |

| Mentions | | | |
|---|---|---|---|
| Countries | -0.0042 | 0.026 | -0.158 |
| Max. Followers | -0.0056 | 0.006 | -0.956 |
| Profession on Twitter | -0.0040 | 0.015 | -0.275 |
| Academic Status | -0.0002 | 0.002 | -0.106 |
| Post Length | 0.0002 | 0.000 | 1.351 |
| HashTags | 0.0061 | 0.024 | 0.251 |
| Twitter Mentions | -0.0222 | 0.030 | -0.748 |
| Author Count | 0.0110 | 0.013 | 0.825 |

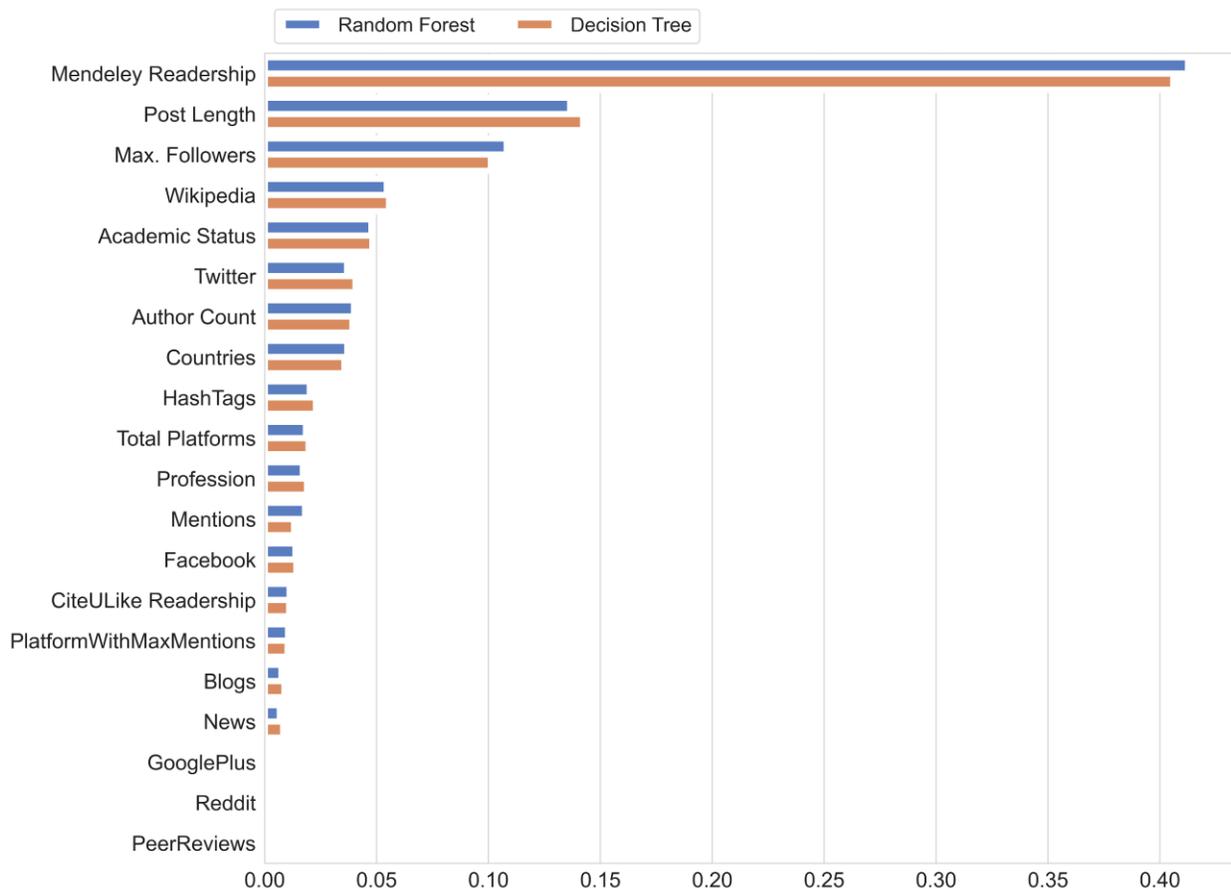

**Figure 5:** The most important features for the Random Forest and Decision Tree models for the 2017 citations in Experiment 3.

## 6. Discussion

Early accurate identification of promising scholarly publications is a great benefit to stakeholders, providing them with the tools to quickly identify important results and act on them sooner than is presently the case. Predicting the number of citations a publication is likely to receive is an essential step toward making this possible. Prediction models that rely on altmetrics are of great interest in this regard as these accumulate just days after a research article has been published. We performed three experiments and found that altmetrics can help in modeling citation prediction. Previous studies on altmetrics focused on identifying correlations with citations and most of the results showed low to medium correlations (Alhoori & Furuta, 2014). However, by implementing machine learning algorithms, we were able to find a non-linear relationship between altmetrics and citations.

We performed citation prediction by dividing the issue into three experiments. Of these three experiments, two involved classification problems and one was a regression problem. In Experiment 1, in which we predicted whether or not an article would receive at least one citation, accuracy was relatively high in comparison with Experiment 2, in which we predicted whether or not an article would receive more or fewer than the median number of citations we had observed in our set of articles. Tree-based and ensemble models performed well in the first experiment, achieving a high level of predictive accuracy and F-1 scores. These models showed that Maximum Followers, Post Length, and Mendeley Readership are some of the most important features for classification. In Expertiment 2, we found that the models were not as accurate as those in Experiment 1. Overall, our models slightly outperformed those presented in Zoller et al. (2016). We found that Mendeley Readership, Maximum Followers on Twitter, and Academic Status are the top three contributors of predictions.

In Experiment 3, we built regression models and achieved an MSE value of 1.50, an MAE value of 0.94, and an R-squared value of 0.435 using the Multiple Linear Regression model. Our regression results are in line with those achieved by Thelwall (2018) and Thelwall and Nevill (2018) in that Mendeley readership was found to be a major factor in predicting eventual citation counts. However, whereas Scopus citations were counted in those studies, we used Google Scholar citations. We also introduced some new features to the prediction, such as Academic Status, Post Length, and Maximum Followers on Twitter. Along with parametric linear models, we also considered non-parametric models such as KNN, SVM, Decision Tree, and Neural Networks in our experiments. We found Mendeley Readership to be the most critical feature across all our models.

Other important factors are Academic Status and Maximum Followers on Twitter. Knowing who are the readers of a publication, plays a vital role in deciding its citation count. For example, compared with undergraduate students, postdocs may be more likely to cite an article from their Mendeley libraries. Further, some important new features include the number of mentions on Wikipedia and the number of mentions on Twitter. Also, the number of countries in which a publication has been shared online. Articles shared online from certain countries may have a higher chance of gaining future citations than is the case for articles shared online in other countries. We plan to investigate these features further in the future.

We have observed that our models in Experiment 2 and mainly Experiment 3 can explain the variance in the short term (i.e., the 2017 citations) but have less explanatory power in the long term (i.e., the 2020 citations). The results shown in Tables 7, 8, 10, and 11 suggest that altmetrics are good predictors of the short-term but not the long-term impact of articles. Popular research articles on social media won't necessarily be popular within the scholarly community. However, this is not the case with articles on Mendeley, which is both widely and predominantly used by researchers, some of whom read and/or cite articles from their Mendeley library. Therefore, Mendeley readership can be considered an essential factor for citation prediction and could provide some useful early signals of the scholarly impact of a publication.

We also collected some articles published in 2013, 2014, and 2016. We used altmetrics from 2016 and citations from 2017 and 2020. We have observed a similar pattern for articles published in 2013 and 2014 to that for articles published in 2015 (i.e., the present study). The pattern does not hold, however, in regard to experiments in which machine learning models are used to predict the scholarly impact of articles published in 2016. Models such as Random Forest, Decision Tree, and AdaBoost showed a decline in F-1 scores as the year of publication approached the year in which early citations were collected. We think some time must pass before altmetrics from multiple platforms accrue for scholarly articles—although less time than is required for traditional citations to accumulate. However, we expect that when collected, altmetrics may be capable of explaining short-term scholarly impact. The length of time that must elapse for altmetrics to accumulate and to function as a useful predictor remains a topic for future research.

There are several ways in which additional research could improve on and advance the direction of the present study. We think the predictive accuracy of the approach to measuring scholarly impact that we have described could be improved by incorporating some traditional features such as publication venue (Alhoori & Furuta, 2017) or h-index. These features have proven effective in other studies. Further, whereas we included a broad array of subject areas in the present study (Figure 1), it may be helpful to consider only individual disciplines in predicting citation counts. It may be that differences across fields account to some extent for the unpredictability we found in relation to modeling citations. It may also be useful to collect citation counts across several platforms (e.g., Google Scholar, Scopus, Dimensions), as it is reasonable to expect the numbers captured to differ from one platform to another. Similarly, it may also be beneficial to use several sources for altmetrics, as some may yield information that is lacking in other sources. It may also prove fruitful to consider more granular detail in relation to the responses that articles receive across online platforms. For example, in some studies, researchers have examined the nature of user responses on Facebook (Freeman et al., 2019, 2020) and Twitter (Sahni et al., 2017). Lastly, our results could be compared with those produced by models that use other predictors of future citation counts, such as early citation counts and journal impact factor, as these additional features have shown promising results in other studies.

## 7. Conclusion

Given the proliferation of research publications, predicting the scholarly impact of research at an early stage would save time for the scholarly community, research agencies, and policymakers and thereby accelerate the overall progress of research. In the present study, we investigated the possibility of using mainly social media features to predict the early and long-term citation counts of research publications. We built and tested several classifiers and regressors using altmetrics data and found that neural networks and ensemble models performed better than the other models in terms of F-1 scores. We also found that in addition to many other factors Mendeley readership plays a crucial role in determining early citations. In the future, we intend to predict the long-term impact of research and to explore more features in this context.

**Acknowledgments**
This work is supported in part by NSF Grant No. 2022443.

odt=0,5

Timilsina, M., Davis, B., Taylor, M., & Hayes, C. (2017). Predicting citations from mainstream news, weblogs and discussion forums. *Proceedings of the International Conference on Web Intelligence*, 237–244.

Tonia, T., Van Oyen, H., Berger, A., Schindler, C., & Künzli, N. (2016). If I tweet will you cite? The effect of social media exposure of articles on downloads and citations. *International Journal of Public Health*, *61*(4), 513–520.

Totti, L. C., Mitra, P., Ouzzani, M., & Zaki, M. J. (2016). A Query-oriented Approach for Relevance in Citation Networks. *Proceedings of the 25th International Conference Companion on World Wide Web*, 401–406.

Valenzuela, M., Ha, V., & Etzioni, O. (2015). Identifying Meaningful Citations. *AAAI Workshop: Scholarly Big Data*. http://www.aaai.org/ocs/index.php/WS/AAAIW15/paper/download/10185/10244

van Dijk, D., Manor, O., & Carey, L. B. (2014). Publication metrics and success on the academic job market. *Current Biology: CB*, *24*(11), R516–R517.

Van Noorden, R. (2014). Global scientific output doubles every nine years. *Nature News Blog*.

Waltman, L. (2016). A review of the literature on citation impact indicators. *Journal of Informetrics*, *10*(2), 365–391.

Wang, D., Song, C., & Barabási, A.-L. (2013). Quantifying long-term scientific impact. *Science*, *342*(6154), 127–132.

Wang, M., Yu, G., Xu, J., He, H., Yu, D., & An, S. (2012). Development a case-based classifier for predicting highly cited papers. *Journal of Informetrics*, *6*(4), 586–599.

Weihs, L., & Etzioni, O. (2017). Learning to Predict Citation-Based Impact Measures. *2017 ACM/IEEE Joint Conference on Digital Libraries (JCDL)*, 1–10.

Willmott, C. J., & Matsuura, K. (2005). Advantages of the mean absolute error (MAE) over the root mean square error (RMSE) in assessing average model performance. *Climate*
33